\documentclass[aps, prl, showpacs, twocolumn]{revtex4}
\usepackage{graphics}
\usepackage{graphicx}
\usepackage{amssymb}
\usepackage{amsmath}
\usepackage{bm}

\newcommand{\bra}[1]{\langle #1 \vert}
\newcommand{\ket}[1]{\vert #1 \rangle}

\begin{document}

\title{Quantum computing with a single molecular ensemble and a Cooper pair box}
\author{Karl Tordrup}
\author{Klaus M\o lmer}
\affiliation{Lundbeck Foundation Theoretical Center for Quantum System Research, Department of Physics and Astronomy, University of Aarhus,
DK-8000 Aarhus C, Denmark}

\date{\today}

\begin{abstract}
We propose to encode quantum information in rotational
excitations in a molecular ensemble. Using a stripline cavity
field for quantum state transfer between the molecular
ensemble and a Cooper pair box two-level system, our proposal
offers a linear scaling of the number of qubits in our
register with the number of rotationally excited states
available in the molecules.
\end{abstract}


\maketitle

\begin{figure}
\includegraphics{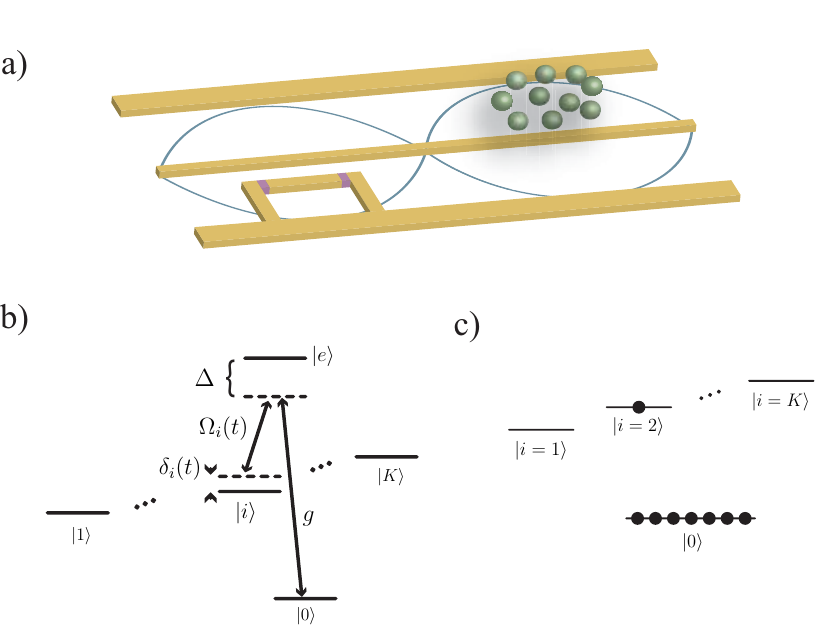}

\caption{(Color online). (a) A Cooper pair box is strongly
coupled to a stripline cavity as realised in \cite{YaleCPB}.
Additionally a cloud of cold polar molecules interacts with
the quantized cavity field. (b) Internal level scheme for a
single molecule. The reservoir state $\ket{0}$ is selectively
coupled to each of the excited states $\ket{i}$ by Raman
transitions. (c) Encoding of $K$ qubits in the symmetric
states of an ensemble of $N>K$ identical
particles.}\label{fig:setup}
\end{figure}

One obstacle which transcends all implementations of a
quantum computer \cite{DeutschJozsaExp, ShorExp, oneWayExp,
couplingCPBs} concerns the extension of current
proof-of-principle operations beyond a handful of qubits. At
the heart of this obstacle lies the exponential scaling of
the Hilbert space dimension with the number of qubits. If one
chooses to work with qubits encoded in separate particles,
the available state space is exponentially large in the
particle number, but selective access to individual particles
and precise control of the interactions among individual
quantum particles presents a formidable challenge. This has
spurred interest in quantum systems that intrinsically
support a vast Hilbert space. Obvious candidates are
molecular quantum systems which easily provide 100 accessible
internal rotational and vibrational levels
\cite{Kosloff,decoherence}. The quantum information capacity
of such systems corresponds, however, to a mere
$\log_2(100)\approx 6$ qubits, and most molecular
implementations to date have not exploited the rich internal
structure, but have focussed on other advantages provided by
molecular systems such as the large intermolecular
dipole-dipole coupling \cite{DeMille}, switchable
interactions \cite{switch}, and long coherence times
\cite{decoherence}. These advantages also make molecules very
attractive for hybrid quantum computing schemes involving
solid state, optical and molecular quantum degrees of freedom
simultaneously. Notably, in \cite{Hybrid}, it has been
proposed to trap a mesoscopic molecular ensemble at an
antinode of the quantized field of a stripline cavity with a
Cooper pair box (CPB) placed at the adjacent antinode. This
setup is illustrated in Fig.\ \ref{fig:setup}a. The energy
scale for the stripline cavity mode matches typical energies
for rotational excitations of polar molecules, providing a
natural interface between the cavity and molecular degrees of
freedom. The large electric dipole moment of polar molecules
makes the strong coupling regime relatively easy to achieve
while strong coupling of the field to the CPB has been
demonstrated experimentally in \cite{YaleCPB, CPBphotonNo}.
Furthermore, by using an ensemble of $N$ molecules one
achieves a $\sqrt{N}$ enhancement of the coupling to the weak
quantum field compared to the single molecule vacuum Rabi
frequency $g$. In \cite{Hybrid}, the essential idea is to
counteract the rapid decoherence in a Cooper-pair box by
transferring the quantum state to the molecular ensemble for
storage of the qubit in a collective molecular excitation
between quantum gates.

In this paper we shall present a method for many-qubit
quantum computing with a single molecular ensemble and a
Cooper pair box. We shall apply an ensemble of $N$ cold polar
molecules with a ground state and $K$ accessible excited
states as illustrated in Fig.\ \ref{fig:setup}b. The
potentially available Hilbert space for the molecular system
is of dimension $(K+1)^N$, but, by limiting ourselves to the
symmetric states with at most one molecule populating each of
the excited states, this is reduced to $2^K$, the Hilbert
space dimension of a $K$-qubit register.

We assume all molecules are initially prepared in the ground
state $\ket{0}$, which is coupled to the excited states
$\ket{i}$ through a Raman process involving the cavity field
coupling constant $g$ and a classical field $\Omega_i(t)$.
Since both fields couple symmetrically to all molecules in
the ensemble, elementary excitations produce the  symmetric
(Dicke) states $\ket{i} = (1/\sqrt{N})\sum_j\ket{0_1 0_2
\ldots i_j \ldots 0_N}$ and so forth, where the index $j$
runs over all molecules in the cloud. We define the
collective raising operator $m_i^{\dag} = (1/\sqrt{N}) \sum_j
\ket{i}_{jj}\bra{0}$. In the regime with only few excited
molecules, the collective operators approximately obey the
bosonic commutator relation $[m_i,m_i^{\dag}]\approx1$ and
the cloud can be treated as a collection of $K$ uncoupled
harmonic oscillators.

The conventional approach to encoding qubits in atoms and
molecules is to encode a single qubit in a single particle
\cite{singleAtoms,DeMille}, or in a single collective degree
of freedom in an atomic or molecular cloud
\cite{ensembleAtoms,Hybrid}. However, as was recently
proposed in \cite{encoding}, one can encode $K$ qubits in a
single cloud of identical particles, each with $K+1$
accessible internal levels by associating the logical
register state $\ket{a_1 a_2 \cdots a_K}$ ($a_i = 0,1$) with
the collective state $\Pi_i (m^{\dag}_i)^{a_i}\ket{0_1 0_2
\ldots 0_N}$ with $a_i$ particles populating the
$i^{\text{th}}$ level, illustrated in Fig.\
\ref{fig:setup}(c). The permutation symmetry among particles
is important, and, for instance, the logical two-qubit state
$\ket{01}$ is an entangled state with no particles populating
the excited state $\ket{i=1}$ and a unit population of state
$\ket{i=2}$ evenly distributed over all the molecules in the
ensemble, in contrast to the conventional simple product
state encoding of the same state. The advantage of this
encoding is that it circumvents the need for addressing of
individual particles, since qubit access is granted by
selective coupling of the ground state to one of the $K$
excited states. The complication of the method lies in the
restriction of the dynamics to the specified state space with
at most one particle populating each of the excited states,
and in the operations on these states that depend on the
population of the other excited states. In \cite{encoding} it
was proposed to use the Rydberg blockade mechanism for
controlled dynamics of neutral atomic ensembles. In this
Rapid Communication we shall describe how the cavity field
and the two-level system offered by the Cooper pair box can
be used to achieve the same goal for polar molecules.

We now turn to the setup indicated in Fig.\
\ref{fig:setup}(a). The Cooper pair box is a superconducting
circuit with an island onto which charge may tunnel through
an insulating barrier as described by a phenomenogical
Hamiltonian, $H=-E_J/2\sum_n
\ket{n}\bra{n+1}+\ket{n+1}\bra{n}$ \cite{CPB}. Due to the
quadratic nature of the electrostatic interaction the energy
levels are non-equidistant, and using resonant transitions
only, the system may at cryogenic temperatures be restricted
to the two lowest quantum states with corresponding raising
and lowering operators, $\sigma^+$ and $\sigma^- $. The CPB
is mounted on a superconducting stripline cavity which can
hold a cavity field with creation and annihilation operators
$c^{\dag}$ and $c$ with very modest field damping
\cite{strongCQED}. The combined system CPB-cavity system is
governed by the Jaynes-Cummings type Hamiltonian
\begin{equation}\label{eq:HCPB}
H_{\text{CPB}} = g_c (\sigma^- c^{\dag} + \sigma^+ c) + \delta_{\text{CPB}}(t)\sigma^+ \sigma^-,
\end{equation}
where
$\delta_{\text{CPB}}(t)=\omega_{\text{CPB}}(t)-\omega_c$ is
the tunable CPB detuning with respect to the cavity field. A
number of phenomena related to the Jaynes-Cummings
Hamiltonian in quantum optics have been observed in the
CPB-cavity system \cite{YaleCPB,CPBphotonNo}, and two-qubit
gates on two Cooper pair boxes coupled to a single cavity
field mode have recently been demonstrated
\cite{couplingCPBs}. The molecular ensemble is addressed by a
Raman transition involving the cavity field and a classical
field with tunable frequency and real amplitude $\Omega_i(t)$
[see Fig.\ \ref{fig:setup}(b)]. In the rotating wave
approximation after adiabatic elimination of the excited
state $\ket{e}$ the coupling of the cavity field to the
$i$'th molecular qubit is described by
\begin{equation}\label{eq:Hm}
H_{\text{M}} =  g_i(t) (m_i c^{\dag} + m_i^{\dag} c) + \delta_i(t)m_i^{\dag}m_i.
\end{equation}
Here $g_i(t) = \Omega_i(t) g \sqrt{N_0}/2\Delta$ is the
effective coupling strength with $\Delta$ the detuning with
respect to the intermediate excited state, and $\delta_i(t)$
is the two-photon Raman detuning of level $\ket{i}$, cf.
Fig.\ \ref{fig:setup}(b). Coupling by higher order Raman
processes with the cavity field and multiple classical field
components allows exploration of a wider range of molecular
states for which Eq.\ (\ref{eq:Hm}) applies with modified
expressions for $g_i(t)$.

We will now describe how quantum information can be encoded
and processed in the combined system. Initially the molecular
ensemble is prepared with all molecules in the zero state
corresponding to all qubits set to the value $0$. We now need
to specify how to carry out reliable one- and two-bit gates
on the system, and specifically for the ensemble encoding, we
have to ascertain that no register state is populated by more
than a single molecule. The cavity and CPB are also prepared
in their ground states, and SWAP operations of arbitrary
unknown states between any qubit component of the molecular
memory and the CPB via the cavity field, combined with an
arbitrary single qubit rotation of the CPB two-level system
by resonant driving with a classical field, implements this
rotation on the desired single qubit of the register. As a
fully entangling two-qubit gate we propose to SWAP one
molecular ensemble qubit to the CPB and then SWAP another
molecular ensemble qubit to the cavity field. The CPB-cavity
interaction can then provide a state dependent phase,
remaining with the two-qubit states when they are finally
returned to the collective molecular ensemble states. Before
presenting details of these processes we note that the main
elements of these steps are, indeed, very similar to the
ideas for quantum computing with atoms coupled via a cavity
field \cite{atomCavity} and with trapped ions, coupled to
each other via their collective motional degree of freedom
\cite{ionTrap}. But we emphasize the significant difference
that our molecules do not need to be individually addressed,
and that we need to pass the cavity excitation through the
Cooper pair box to restrict the Hilbert space to two states
per qubit degree of freedom and to provide the interaction in
the system.

\begin{figure}
\includegraphics[width=8.6cm]{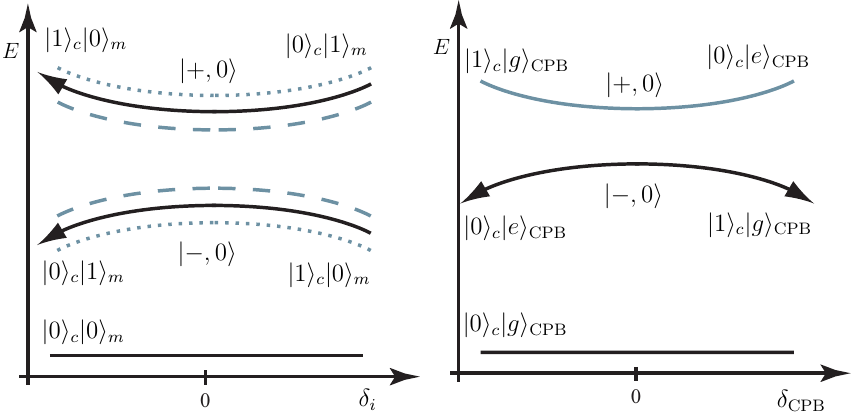}
\caption{Left: In order to transfer a molecular state to the
cavity we turn on a far-detuned field with $\delta_i /g_i \gg
1$ and sweep through resonance adiabatically following the
dressed state $\ket{+,0}$. To transfer the state back to the
molecular ensemble the field is again turned on at $\delta_i
/ g_i \gg 1$, this time following the $\ket{-,0}$ state
through resonance. Phase errors due to fluctuating $N_0$
(illustrated by the dashed and dotted lines) are exactly
cancelled leaving only a geometric phase of $e^{-i \pi}$.
Right: When transferring states between the cavity and CPB
one must follow the same path back and forth hence the
geometric phase vanishes. Since there is no fluctuation in
the interaction strength $g_c$ the dynamical phase can be
tailored to 0 mod $2\pi$. }\label{fig:dressed}
\end{figure}

The SWAP operations can be realized by adiabatically sweeping
the detunings across resonance. When transferring a molecular
state to the empty cavity the coupling $g_i(t)$ is turned on
with $\delta_i(t=0)/g_i \gg 1$. As $\delta_i(t)$ passes
through resonance each basis state adiabatically evolves into
the corresponding dressed state
\begin{eqnarray}
\ket{+,n} &=& \cos (\theta)\ket{1}_m\ket{n}_c+\sin (\theta)\ket{0}_m\ket{n+1}_c \\
\ket{-,n} &=& -\sin (\theta)\ket{1}_m\ket{n}_c+\cos (\theta)\ket{0}_m\ket{n+1}_c
\end{eqnarray}
with $\tan(2\theta) = 2g_i \sqrt{n+1}/\delta_i(t)$ and
energies $E_{\pm ,n} = \tfrac{1}{2}\delta_i(t)
\pm\tfrac{1}{2}\sqrt{\delta_i^2(t)+4g_i^2(n+1)}$. Here $n+1$
is the number of elementary excitations i.e. $n=0$ for the
states $\ket{0}_m\ket{1}_c$ and $\ket{1}_m\ket{0}_c$, and
$n=1$ for $\ket{1}_m\ket{1}_c$. As a result each state
acquires a nonlinear phase relative to the ground state of
$\varphi_{\pm,n} = \tfrac{1}{2}\int_0^T -\delta_i(t)\mp
\sqrt{\delta_i^2(t)+4g_i^2(n+1)}dt$. Thus the logical state
$\ket{1}$ acquires a phase relative to $\ket{0}$ while
following $\ket{+,0}$ (see Fig.\ \ref{fig:dressed}, left
panel). When returning the state to the ensemble the
$\ket{1}$-component can either be made to follow $\ket{+,0}$
(by sweeping from negative to positive detuning) or
$\ket{-,0}$ (by sweeping from positive to negative detuning).
By choosing the latter option the total dynamical phase
$\text{exp}\{-i\int_0^T [E_+(t)+E_-(t)]dt\}$ is exactly
cancelled for any sweep that is antisymmetric about $t=T/2$.
The $\ket{1}$ state then only acquires a geometric phase of
$e^{-i\pi}$. The effective coupling of the ground to any
particular excited states is amplified by the number of
participating molecules, i.e., it is $g\sqrt{N_0}$ with $N_0$
the number of ground state molecules. This number has a
quantum mechanical uncertainty when some of the other excited
states are in superposition states of being populated and
unpopulated. The relative variation in the coupling strength
due to this uncertainty can be suppressed by requiring $N\gg
K$, but we observe that since we accumulate opposite phases
in the two adiabatically swept passages, indicated by the
dotted ($N>N_0$) and dashed ($N<N_0$) curves in the figure,
the difference will effectively cancel.

The phase dynamics is most easily visualized by viewing the
dynamics as a spin-1/2 precessing about a fictitious magnetic
field $\bm{B}(t) = -(g_i,0,\tfrac{1}{2}\delta_i(t))$. Two
consecutive sweeps of $\delta_i(t)$ in the same direction
corresponds to a $2\pi$ rotation of $\bm{B}(t)$ giving a
geometric phase of $e^{-i\pi}$. This SWAP operation thus
incurs an extra single qubit Z-gate which must be absorbed
into future operations. When transferring between the cavity
and CPB we have an \emph{always on} interaction $g_c$ and two
consecutive sweeps must be in opposite directions (see Fig.\
\ref{fig:dressed}, right panel). This corresponds to
following the $\ket{+,0}$ ($\ket{-,0}$) state twice and hence
there is no cancellation of the dynamical phase. However in
this case the geometric phase vanishes since the fictitious
B-field traces out a path enclosing a vanishing solid angle.
The sweep may then be chosen such that the dynamical phase
becomes 0 mod $2\pi$ since there is no fluctuation in the
coupling $g_c$. For instance the cubic parametrization
$\delta_{\text{CPB}}(t) = \delta_0 (2t/T-1)^3$ with
$\delta_0=19.24g_c$ and $T=20.77g_c^{-1}$ gives cancellation
of the dynamical phase. The cubic form was chosen since it is
odd about $t=T/2$ and offers faster implementation than a
linear chirp while remaining in the adiabatic limit.

\begin{figure}
\includegraphics[width=5.1cm]{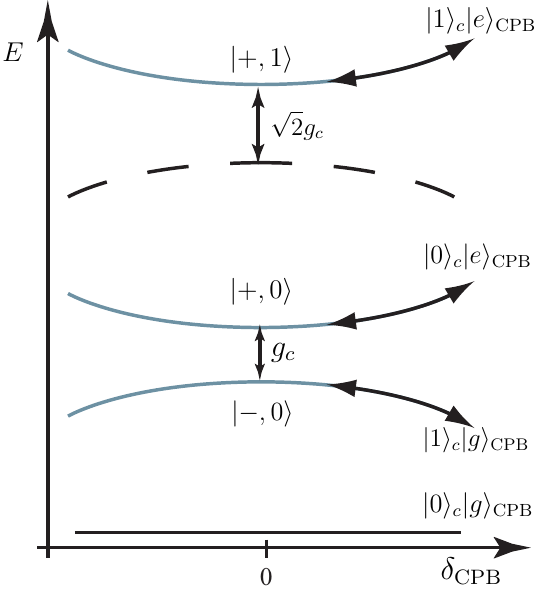}
\caption{A conditional phase shift is performed by exploiting that the dressed state energies are
nonlinear in the excitation number $n$. Starting at $\delta_{\text{CPB}}(0)/g_c \gg 1$ the CPB is
tuned close to resonance ($\delta_{\text{CPB}}(T/2) \sim g_c $) and back to end at
$\delta_{\text{CPB}}(T)/g_c \gg 1$. The functional form of $\delta_{\text{CPB}}(t)$ is chosen such
that $\varphi_{+,1}=\pi$ mod $2\pi$ while $\varphi_{\pm,0} = 0$ mod $2\pi$. The dashed line represents
the dressed state $\ket{-,1}$ which at $\delta_{\text{CPB}}/g_c \gg 1$ corresponds to $\ket{2}_c
\ket{g}_{\text{CPB}}$ i.e. a doubly excited state which is not part of the computational Hilbert
space.}\label{fig:Cphase}
\end{figure}

To implement a single qubit gate on qubit $i$ we transfer the
molecular state to the cavity by tuning $\delta_i(t)$ across
resonance and then transfer the state to the CPB by a similar
sweep of $\delta_{\text{CPB}}(t)$. A single qubit gate may
then be implemented on the CPB using microwave pulses
\cite{cpbQI} whereafter the state is transferred back to the
ensemble via the cavity. To implement two-qubit entangling
operations we make use of the cavity-CPB coupling described
by Eq.\ (\ref{eq:HCPB}). The control qubit is transferred to
the CPB and the target qubit is subsequently transferred to
the cavity field as described above. From an initial detuning
of $\delta_{\text{CPB}}/g_c \gg 1$ the CPB detuning is tuned
close to resonance and back, the computational states
evolving adiabatically along the dressed states as seen in
Fig.\ \ref{fig:Cphase}. In order to implement a fully
entangling controlled phase operation the functional form of
$\delta_{\text{CPB}}(t)$ must be chosen such that
$\varphi_{\pm,0} = 0$ mod $2\pi$ and $\varphi_{+,1} = \pi$
mod $2\pi$. We find that parametrizing
$\delta_{\text{CPB}}(t) = a(2t/T-1)^2+b$ with $a=33.05g_c$,
$b=0.6664g_c$ and $T=58.07/g_c$ produces a controlled phase
gate with near unit fidelity.

Let us now address the feasibility of our proposal with
current physical parameters. For the Cooper pair box, the
dominant source of decoherence is second order charge noise,
which is minimized by operation at the so-called sweet spot
leading to dephasing times of the order $T_2 \sim 1$ $\mu$s
\cite{transmon}. By comparison the vacuum Rabi frequency of
the CPB-cavity field coupling is of the order $g_c \sim 2\pi
\times 50$ MHz, so for single qubit rotations on the CPB we
obtain $g_c T_2 \sim 300$; i.e., one can implement on the
order of 300 CPB-cavity SWAP operations before the qubit
decoheres. The conditional phase gate we have proposed has a
gate time of $T = 58.07g_c^{-1}$ giving $T/T_2 \sim 0.1$.
This ratio can be improved by more than an order of magnitude
by replacing the conventional CPB with a recently improved
so-called transmon design \cite{transmon}, which operates in
a regime where the dominant decoherence process is relaxation
with $T_1 \sim 16$ $\mu$s. The coherence time of the
molecular ensemble is limited mainly by the collision rate
and by coupling due to the long-range dipole-dipole
interaction. In a magnetic trap,  at $T=1$ mK the scattering
rate due to the asymptotic $r^{-6}$ interaction is estimated
to be $\gamma \lesssim 2\pi \times 700$ Hz \cite{Hybrid}. In
an electrostatic trap the induced dipole moment
$\mu_{\text{ind}}$ leads to an $r^{-3}$ interaction, but at
$T=1$ mK and $\mu_{\text{ind}} < 1$ D, the above estimate for
the scattering rate still holds and we conclude that the
interactions within the molecular ensemble should not
significantly lower the effective decoherence time. With
photon loss rates down to $2\pi \times 10$ kHz and realistic
values of $g_i$ up to $2\pi \times 10$ MHz \cite{Hybrid},
with current technology one could implement hundreds of
gates, sufficient to provide proof of concept for the present
scheme and to carry out simple error correction algorithms.

In conclusion we have described a system which with current
technology could provide quantum computation with an
appreciable number of qubits. The ensemble encoding we have
used is applicable to systems which contain a saturable
element such as the Cooper pair box which furthermore serves
to provide an entangling conditional phase gate. The
molecular ensemble provides an efficient encoding scheme for
many easily accessible qubits and long coherence times which
may be improved even further by adopting a crystalline
ensemble in which collisions are suppressed. If additionally
one chooses a trapping scheme which is independent of the
rotational state, there exist certain "magic" configurations
for which detrimental coupling to the phonon spectrum is
completely suppressed \cite{crystalMemory}. Potentially in
future applications several ensembles could be coupled to an
array of Cooper pair boxes in several interconnected cavities
providing scalability as well as the benefit of parallel
processing. Further improvements on the present work could
include the application of Optimal Control Theory
\cite{decoherence} to improve the gate time for the
conditional phase shift as well for the SWAP operations.

This work was supported by the European Commission through
the Integrated Project FET/QIPC "SCALA".

\bibliography{minbib}
\end{document}